
%
%
\input phyzzx

\overfullrule=0pt
\font\twelvebf=cmbx12
\nopagenumbers
\footline={\ifnum\pageno>1\hfil\folio\hfil\else\hfil\fi}
\line{\hfil March, 1994}
\line{\hfil CU-TP-628}
\line{\hfil IASSNS-HEP-93/34}

\vglue .4in
\centerline {\twelvebf Quantum Tunnelings with Global Charge}
\vskip .5in
\centerline{\it  Kimyeong Lee$^\dagger$ }

\vskip .1in
\centerline {Physics Department, Columbia University}
\centerline {New York, New York 10027}
\vskip .2in

\vskip .5in
\centerline {\bf Abstract}
\vskip .1in

We investigate quantum tunneling in the theory of a complex scalar
field with a global $U(1)$ symmetry when the charge density of the
initial configuration does not vanish.  We discuss the possible final
configurations and set up the Euclidean path integral formalism to
find the bubble nucleation and to study the bubble evolution. For the
stationary path, or the bounce solution, in the Euclidean time, the
phase variable becomes pure imaginary so that the charge density
remains real.  We apply this formalism to examples when the initial
charge density is small. While the phase transition considered here
occurs in zero temperature, the bubble dynamics is richly complicated,
involving conserved charge, the sound wave and the supersonic bubble
wall.

\vfill

\footnote{}{$^\dagger$ Email Address: klee$@$cuphyf.phys.columbia.edu}

\vfill\eject

\def\pr#1#2#3{Phys. Rev. {\bf D#1}, #2 (19#3)}
\def\prl#1#2#3{Phys. Rev. Lett. {\bf #1}, #2 (19#3)}

\def\np#1#2#3{Nucl. Phys. {\bf B#1}, #2 (19#3)}
\def\pl#1#2#3{Phys. Lett. {\bf #1B}, #2 (19#3)}

\REF\rBernstein{ J. Bernstein and S. Dodelson, \prl{66}{683}{91}; K.M. Benson,
J. Bernstein and S. Dodelson, \pr{44}{2480}{91}; K.M. Benson and L.M.
Widrow, \np{353}{187}{91}.}

\REF\rKlee{K. Lee, \prl{61}{263}{88};
S. Coleman and K. Lee, \np{329}{387}{90}; K. Lee, \pr{48}{2493}{93}.}

\REF\rColeman{S. Coleman,  \pr{15}{2929}{77}; C.G.
 Callan and S. Coleman, \pr{16}{1762}{77}; S. Coleman, \np{307}{867}{88}.}

\REF\rTdlee{ T.D. Lee, \pr{10}{2739}{76}; S. Coleman \np{262}{263}{85};
T.D. Lee and Y. Pang, Phys. Rep. {\bf 221}, 251 (1992).}

\REF\rSpector{ D. Spector, \pl{194}{103}{87}.}



\chapter{Introduction}

Recently, there have been some interests in the first order phase
transition involving nonzero global charge. The finite temperature
effective potential and the phase structure in this type of the phase
transition have been extensively studied.\refmark\rBernstein However,
how the phase transition proceeds has not been discussed in detail.
Here we investigate the phase transition in a model which involves
nonzero charge and a nontrivial bubble wall dynamics. This model is
the theory of a complex scalar field with a global abelian symmetry.
Even in the zero temperature phase transition, this model has a rather
rich dynamics in the phase transition, depending on the initial
configuration and the potential energy.  Since there are only two
field degrees of freedom in this model, this model can be rather
easily approached analytically and numerically. We hope that our toy
model illuminates some aspects of the phase transition involving
global charge and that some of insights gained would be applicable to
the QCD phase transition and the electroweak phase transition.

The general formalism of the Euclidean phase intergal involving
nonzero charge in our model has been developed sometime
ago.\refmark\rKlee While this formalism has been applied to wormhole
physics, there has been no direct attempt to apply for the first order
phase transition.  We extend this formalism to the first order phase
phase transition at zero temperature, following the standard
formalism.\refmark\rColeman One interesting aspect of our formalism is
that the stationary path of the angle variable of the complex scalar
field becomes pure imaginary. From this Euclidean path integral, one
can find the bounce solution and calculate the bubble nucleation rate.
In addition, one can gain some insights on the bubble evolution.

The initial configuration we are interested in here is a homogeneous
configuration which is classically stable but not quantum
mechanically. The charge density of the initial configuration is
nonzero and uniform. The final configuration after the phase
transition, it turns out, could be more complicated than the
configuration of the lowest potential energy.  In our model, there
could be an attractive force between charges and charges clump
together forming Q-balls.\refmark\rTdlee Thus, the final configuration
could be inhomogeneous with Q-balls floating in the symmetric phase.

The Euclidean path integral allows us to calculate the imaginary part
of the energy for the metastable initial configuration by the
semiclassical method. The contribution to the path integral is
dominated by bounce solutions. Contrasted to the usual
case,\refmark\rColeman the initial charge density now breaks the
$O(4)$ symmetry to the $O(3)$ symmetry.  In this paper we focus  the
case of  small initial charge density, where the bounce solution is
close the $O(4)$ bounce solution of zero charge.  When the charge
density is small, we can look at the perturbative correction to the
$O(4)$ symmetric solution. The current conservation equation in this
background turns out to be a boundary value problem in the classical
electrodynamics and can be solved in the thin wall limit. This leads
to some insights on the current flow in the bounce solution and the
deformation of the $O(4)$ symmetric bounce solution.  This in turn
leads to an understanding of the bubble evolution via the analytic
continuation.

When there is nonzero charge density and the initial configuration is
metastable, there is always the sound wave of the speed less than the
speed of light.  When the bubble of a ``true vacuum'' is nucleated, it
will expand.  The bubble wall speed could reach the sound speed of the
initial configuration in finite time, becoming supersonic.  In
addition, some bubbles speed up to the speed of light in finite time,
which implies some sort of a new instability.

The plan of this paper is as follows. In Sec.2, we introduce the
theory of a complex scalar field. We study the stability condition of
the possible initial configurations and discuss the final
configurations we expect after the phase transition.  In Sec.3. we
study the Euclidean path integral.  We offer a formal discussion of
the bubble nucleation and evolution. In Sec.4, we examine in detail
various examples when the initial charge density is small, and present
the qualitative pictures of the bubble nucleation and evolution. In
Sec.5, we conclude with some remarks and questions.

\chapter{Model}

We study the theory of a complex scalar field $\phi =
fe^{i\theta}/\sqrt{2}$ with a global $U(1)$ symmetry. The lagrangian
is given by
$$ \eqalign{ {\cal L} &\ = |\partial_\mu \phi |^2 - U(\sqrt{2} \phi)
\cr
&\ = {1\over 2} (\partial_\mu f)^2 + {1\over 2}f^2 (\partial_\mu
\theta)^2 - U(f) \cr}
\eqno\eq $$
The global symmetry arising from a constant shift of $\theta$ leads to
the conserved current
$$ J_\mu = - i(\phi^* \partial_\mu \phi - \partial_\mu \phi^* \phi)
=  f^2 \partial_\mu \theta    \eqno\eq $$
The total charge is $Q=\int d^3x f^2 \dot{\theta}$. While there could
be global strings in this theory, they seem not to play any essential
role in our discussion and will be neglected here.

\FIG\fone{Plot of $e(\rho, f)$ and $U(f)$ for a typical case.}

In this paper we are interested in quantum evolution of a metastable
initial configuration with nonzero charge density. To start,  we
should have a proper description of possible initial configurations.
While some of them could be inhomogeneous and localized, here
we will focus on the static homogeneous initial configurations
$(f,\theta = wt)$ with a given charge density $\rho = f^2
\dot{\theta}= f^2w$.  The energy density for these configurations is
given by
$$ e(\rho,f) = {\rho^2 \over 2f^2} + U(f)   \eqno\eq $$
The first term in the right hand side of Eq.(2.3) is the centrifugal
term due to the conserved charge. An initial configuration would lie
at the local minimum of $e(\rho,f)$ to be stable under homogeneous
fluctuations, and satisfies
$$ {\partial e\over \partial f}  = - {\rho^2 \over f^3} + U'(f) = 0 \eqno\eq $$
and $ {\partial^2 e \over \partial f^2} > 0 $ or
$$ 3{\rho^2 \over f^4} + U'' > 0  \eqno\eq $$
We denote a  local minimum  by $f_0$ and  $w_0 = \rho /f_0^2$.
Fig.1   shows an example of $e(\rho,f)$ and $U(f)$ for  nonzero  $\rho$.

A metastable initial configuration should be stable, even
under small inhomogeneous fluctuations, $f_0 + \delta f$ and $\theta =
w_0 t + \delta \theta$. From the classical field equation, we get the
linear equation for the fluctuations,
$$ \eqalign{ &\ -\delta \ddot{f} + \partial_i^2 \delta f + w^2_0
\delta f  + 2w_0 f_0 \delta
\dot{\theta} - U''(f_0) \delta f = 0 \cr
&\ - 2w_0 f_0 \delta \dot{f} - f^2_0 \delta
  \ddot{\theta} + f_0^2 \partial_i^2 \delta \theta
=0 \cr }
\eqno\eq $$
With $\delta f , \delta \theta \sim e^{i(\alpha t - \vec{k}\cdot
\vec{r})} $, Eq.(2.6) yields a dispersion relation
$$ [\alpha^2 - \vec{k}^2][\alpha^2 - \vec{k}^2 + w^2_0 - U''(f_0)]
 - 4\alpha^2w_0^2 = 0
\eqno\eq $$
or
$$ \alpha^2 = \vec{k}^2 + {3w_0^2 + U'' \over 2} \pm \left\{ \bigl(
{3w_0^2 + U'' \over 2}  \bigr)^2 + 4w_0^2 \vec{k}^2 \right\}^{1/2}
\eqno\eq $$
One can see that there are one massive and one massless modes. The
massless mode can be interpreted as the sound wave and satisfies the
dispersion relation $\alpha^2 = v_s^2 \vec{k}^2$ for the long wave
length, or for small $\vec{k}$, where the sound speed is given by
$$ v_s^2 = { f_0^4 U''(f_0) - \rho^2 \over f_0^4 U''(f_0) + 3\rho^2 }
\eqno\eq $$
Thus, the configuration is stable under small fluctuations only if
$$ f_0^4 U''(f_0) - \rho^2 > 0 \eqno\eq $$
which is a stronger condition than Eq.(2.5).

When the charge density is high, the centrifugal term would be
balanced by the highest order term in the potential, say $U \approx
\epsilon f^n /n$. From Eqs.(2.4) and (2.9) we see that the sound speed in
this limit is given as
$$ v_s = \sqrt{ n-2 \over n+2} \eqno\eq  $$
It is interesting to note that the potentials of  $n=4$ in four
dimensions,  $n=6$ in three dimensions and $n=\infty$ in two
dimensions are renormalizable interactions and that the corresponding
sound speeds $1/\sqrt{3}, 1/\sqrt{2}, 1$ are those of  hot
relativistic gases in the corresponding dimensions.

The stability condition (2.10) can be examined more concretely in
 the potential
$$ U(f) = {m^2 \over 2}  f^2 + {g  \over 4} f^4 +{\lambda \over 6}f^6
+ {\kappa \over 8}f^8 \eqno\eq $$
(Here we are not concerned about the renormalizability of the theory.
We are interested in the general characteristics of the tunneling when
global charge is involved.)  Assume that the charge density is very
small compare with other scale.  When $m^2 > 0$, the condition (2.4)
can be satisfied for $f_0$ near the symmetric phase,
$$ f_0= \sqrt{\rho/m}+ {\cal O}(\rho^2)
\eqno\eq$$
The stability condition (2.10) becomes
$$ f^4_0 U''(f_0) - \rho^2 = 3g({\rho \over m})^3( 1 + {\cal O }(\rho))
> 0   \eqno\eq $$
which is satisfied only if $g >0$. The configuration is thus stable
only if there is a short range repulsion due to the self interaction.
The velocity of the sound in the case $g>0$ is given as
$$      v_s^2 =   { 3g \rho \over 4m^3} (1 + {\cal O}(\rho))
\eqno\eq $$
which is much smaller than the  unity when the charge density is small.

In the case where the potential (2.12) takes a local minima at $f=v$,
$U \approx {1\over 2} m'^2 (f-v)^2$ with $m'^2 = U''(v)$.  The
condition (2.4) implies $\rho^2 \approx m'^2 v^3(f_0-v)$. The
stability condition (2.10) becomes
$$ m'^2 v^4 +3\rho^2 + v\rho^2
U'''(v)/m'^2 +{\cal O}(\rho^3) > 0
\eqno\eq $$
which is automatically satisfied when
the charge density is small. The sound velocity becomes
$$ v_s^2 = 1 - {4\rho^2 \over m'^2 v^4} + {\cal O}(\rho^4)
\eqno\eq $$
This shows that the massless Goldstone boson becomes the sound wave as
the nonzero charge density is introduced in the broken phase.

We have examined the stability condition on the possible initial
configurations. If an initial configuration is metastable, it will
evolve quantum mechanically so that the potential energy is converted
to the kinetic energy of bubbles and the radiation energy.  The
important question is how we know whether an initial configuration is
metastable quantum mechanically? We would say a configuration is not
stable quantum mechanically if we can find a field configuration of a
lower energy and the same quantum number so that there is no
superselection rule preventing the transition between these two
configurations.

In the case of the theory of a real scalar field, the answer comes
immediately from the potential energy density.\refmark\rColeman The
potential energy density for the final configuration would be the
lowest. A metastable initial configuration can decay via nucleations
of bubbles whose interior is in the true vacuum. We know well how
this phase transition proceeds.

\FIG\ftwo{Plot of  two  potentials $U_1(f)$ and $U_2(f)$.}

For the theory of a complex scalar field, the story is more
complicated as there are two degrees of freedom which work together or
against each other.  If we regularize the system in a large finite
box, we expect that the possible final configuration has the minimum
energy for a given total charge. The excessive energy of the initial
configuration would be channeled into elementary excited modes in the
final configuration through the radiations and the bubble collisions.
In general, it is not easy to find such a final configuration. A
possible final configuration could be inhomogeneous.  In this paper,
we concentrate on the cases where the potential $U(f)$ has a local
minimum at $f=0$ and the phase transition occurs between the $f
\approx 0$ and $f\ne 0$ phases.  Fig.2 shows such potentials. In
addition, we assume that the initial charge density is small compared
with other scales in the problem.

In these limits, we can analyze the problem in a somewhat satisfactory
way.  As far as the $f$ field is concerned, this field wants to settle
at the ground state of $e(\rho, f)$ in Eq.(2.3).  However, we will see
that the charge will  move so that the energy per unit charge
$e(\rho,f)/\rho$ takes the lowest value. Since we concern the phase
transition between the symmetric and asymmetric phases, we choose
$U(f=0)=0$. It turns out that there are three different cases of the
phase transitions to consider: {\it Case A}) The ground state of $U(f)$ is an
asymmetric phase where $U(f)<0$ and the initial configuration is in
the symmetric phase; {\it Case B}) The potential $U(f)$ has
a local minima at an asymmetric phase where $U(f)>0$ and the initial
configuration is in the asymmetric phase; {\it Case C}) The potential
$U(f)$ is identical to  {\it Case B}  but the initial configuration is in
the symmetric phase.  In Fig.2  $U_1(f)$ corresponds to {\it Case A}
and $U_2(f)$ does to {\it Cases B,C}.

Let us first study the first {\it Cases B,C}  where the symmetric phase is
the ground state of $U(f)$. To find out the final configuration, let
us recall that in these cases there could be Q-balls in the theory for
an appropriate potential. Let us here recapitulate the Q-ball
physics briefly.\refmark\rTdlee In the symmetric phase  charged particles
have  mass $m$ and so the ratio $e/\rho$ would be $m$.  For a given
large charge $Q$, the lowest energy state however does not need to be
made of a collection of these charged particles at rest, whose energy
is $m|Q|$. Rather, that  could be a Q-ball if there
is an enough attraction between the charged particles.

\FIG\fthree{ Plot of $  \sqrt{2U(f)/f^2}$ for
three potentials $U_1,U_2,U_3$. }

To find   the condition on the potential for Q-balls to exist,
let us examine a homogeneous configuration whose energy
per unit charge is lowest. We first minimize $e/\rho$ with respect to
$\rho$,
$$ {\partial (e/\rho)   \over \partial \rho}
= {1\over 2f^2} - {1\over \rho^2} U(f) = 0.
\eqno\eq $$
The charge density  is  then fixed as a function of $f$,
$$ \rho = 2f^2 U     \eqno\eq $$
The energy per unit charge is given by
$$ {e \over \rho} = \sqrt{2U \over f^2}
\eqno\eq $$
We have to minize $e/\rho $ in Eq.(2.20) with respect to the $f$
field.  Fig.3 shows $\sqrt{2U/f^2}$ for various potentials.  When
$\sqrt{2U/f^2}$ takes the local minimum value $w_*$ at the nonzero
$f=f_*$ field, such configuration is called Q-matters whose charge
density is fixed to be $\rho_* = f^2_* w_*$.  A Q-ball is a sphere
whose inside is made of Q-matters and outside is just the symmetric
phase.  When the total charge is large, the size of a Q-ball would be
large and the surface energy will be negligible compare with the
volume energy and the energy per charge would be very close to $w_*$
Thus, if $w_*$ is less than $m$, the Q-balls are stable against
decaying into charged particles. Such Q-balls are possible with
$U_1,U_2$ of Fig.3.

For {\it Case B}, the initial configuration should be classically
stable, which means $g>0$ as shown in Eq.(2.15). $U_2$ in Fig.3
represents such a potential.

When the stable Q-balls are possible, the final configuration for {\it
Cases B, C} is inhomogeneous. There will be a region with Q-matters
and the rest as the symmetric phase of zero charge density. The domain
walls separating two regions would slowly evolve to reduce the surface
energy. The ratio of the volumes between two regions would vary
depending on the initial charge density.  If the initial charge
density is small, Q-balls will float in the symmetric vacua. If the
initial charge density is large, the balls of the symmetric vacua will
float in the Q-matter.  In {\it Case B} the phase transition would
proceed with the nucleations of Q-balls and in {\it Case C} it would
proceed with the nucleations of bubbles with less charge density.

If Q-balls are impossible because $w_* >m$ as in $U_3$ of Fig.3, then
the final configuration would be in the symmetric phase with nonzero
charge density. If the initial charge density is larger than $\rho_*$,
the final configuration could be more complicated would not be
discussed here.

To understand the condition (2.18) better, consider
the pressure of a homogeneous configuration given by
$$  p = {\rho^2 \over 2f^2} - U(f)
\eqno\eq $$
We see Eq.(2.18) is identical to the zero pressure condition. When
Q-balls can exist, we have argued that the final configuration could
be inhomogeneous where  Q-balls float in the symmetric vacuum.
The zero pressure condition means that there is no pressure difference
between Q-balls and the symmetric vacuum, leading to an  equilibrium
situation.

Having analyzed the last two cases where the symmetric phase is the
ground state of $U(f)$, let us now consider {\it Case A} where the
asymmetric phase is the ground state of $U(f)$.  For this case, both
the $f$ field and charge prefer the asymmetric phase because the the
energy per charge in the asymmetric phase would be negative compare
with that in the symmetric phase.  The final configuration would be in
the asymmetric phase with uniform charge density. An interesting
observation on this case has been made when the initial state is not
classically stable because $g<0$ in the potential
(2.12).\refmark\rSpector Note that the initial configuration of zero
charge density is classically stable.  When we introduce a small
charge density, charges get  concentrated in several regions and these
regions would evolve classically into the true vacuum.  For the
initial configuration at the symmetric phase to be stable, there
should be a repulsive force between charges $(g>0)$ in long distance.

We note that the energy per unit charge and the pressure are sensitive
to the shift of the potential energy. In the above arguments, $U(0)=0$
was used crucially because we are dealing with tunneling between the
symmetric phase and the asymmetric phase. If we attempted to
understand the tunneling between two asymmetric phases, we would have
needed another device to figure out the final state. We will try to
investigate this example elsewhere.

\chapter{ Quantum Tunneling }

We have studied the general characteristics of the phase transition
from a homogeneous initial configuration with nonzero charge density.
We found out qualitatively what will be the final configuration after
the quantum tunneling. We now want to approach this problem more
analytically by using the Euclidean path integral
formalism.\refmark\rColeman When a nonzero global charge is involved,
the standard formalism should be extended to accommodate the
nontrivial boundary term.\refmark\rKlee Here we summerize and expand
the known results.

We start with the Euclidean generating functional,
$$ <F| e^{-H T }|I> =  \int  [fdf d\theta] \Psi_F^*  e^{- S_E}  \Psi_I
\eqno\eq $$
where the Euclidean action is given by
$$ S_E = \int d^4x \left\{  {1\over 2}  (\partial_\mu f )^2
 + {1\over 2} f^2
(\partial_\mu \theta)^2  + U(f) \right\} \eqno\eq $$
The initial and final states $\Psi_{I,F}$ describe the
configurations of charge density, $\rho_{I,F}$, or
$$  \Psi_{I,F}(\theta, f)  \sim    \exp \left\{  i\int d^3x
\rho_{I,F} \theta \right\}
\eqno\eq $$
In our case, the initial and final states are identical and describe
the initial metastable configuration.

By calculating  the imaginary correction to the energy of the metastable
initial configuration, we can find out the tunneling rate or the bubble
nucleation rate. When we sum over the multibounce contributions to the
energy,
$$ \eqalign{ e^{-{\cal E}VT} &\  = <i|e^{-HT}|i> \cr
&\ = K_0 e^{-{\cal E}_0 VT} \left\{ 1 + VT K  e^{-B} +
{1\over 2} (VTK  e^{-B})^2
+ ....\right\} \cr
&\ = \exp \{ - ( {\cal E}_0 - Ke^{-B}) VT \} \cr}
\eqno\eq $$
where ${\cal E}_0$ is the energy density of the initial configuration,
$K_0$ is the prefactor from the small fluctuations around the initial
configuration, the $B$ factor is the difference between the bounce
action and the background action, $ {\cal E}_O VT$, and $KK_0$ is the
prefactor arising from the small fluctuations around the bounce
solution.  There is a negative mode around the bounce solution which
implies the factor $K$ pure imaginary.  The bubble nucleation rate per
unit volume is then given by $2|K|e^{-B}$.

The path integral will be dominated by the stationary configurations
of the path integral. In our case the boundary condition on the $f$
field of such stationary configuration is fixed to be the initial
classical configuration. However, Eq.(3.3) implies that the boundary
condition on the $\theta$ field is free. Thus, the stationary
configuration in the path integral (3.1) satisfies the Euler equation
for the action $S_E + \Sigma$, where $\Sigma$ is the boundary term
$$ \Sigma = -i \int d^3r  \left\{ \rho_F \theta (\vec{r}, \tau_F) - \rho_I
\theta(\vec{r},\tau_I) \right\}  \eqno\eq $$
{}From the Euclidean equation from the combined action $S_E + \Sigma$,  one can
see
that the stationary path of the angle variable $\theta$ should be pure
imaginary,
$$ \theta \equiv -i\eta  \eqno\eq $$
The Euclidean field equations  for the $f,\eta$ field are  then
$$\eqalign{ &\ \partial^2_\mu f + f (\partial_\mu \eta)^2 - U'(f) =  0  \cr
  &\  \partial_\mu (f^2 \partial_\mu \eta ) = 0 \cr}
\eqno\eq $$
The boundary condition on $\eta$ becomes
$$ f^2 {\partial\eta \over \partial \tau} (\vec{r},\tau_{I,F} ) =
\rho_{I,F}(\vec{r}) \eqno\eq $$
The $f$ field should approach the time independent $f$ given by the
initial configuration. We assume that there is no vortex in the
initial configuration. Then the classically  stable configuration for
given charge should satisfy
$$ \partial_i^2 f + {\rho^2 \over f^3} - U'(f) = 0
\eqno\eq $$
since $\partial_\tau f =0 $ and $\partial_i \eta$ at the boundary.  In
our case, the initial configuration is homogeneous in space and so
Eq.(3.9) becomes identical to Eq.(2.4).

The solution of Eq(3.7) is the so-called  bounce for quantum tunneling.
When the initial charge density vanishes, we know that the $O(4)$
symmetric solution of Eq.(3.7) has been shown to exist by the
undershoot-overshoot method.\refmark\rColeman In our case, the
boundary condition reduces this $O(4)$ symmetry to the $O(3)$ symmetry
because the charge density selects a preferred time direction.   Thus,
we are interested here the bounce which is O(3) symmetric invariant
under the spatial rotation.  This makes Eq.(3.7) a partial
differential equation, whose solution is much harder to find. We can
either use some analytic tools or numerical analysis to find the
bounce solution. In the next section, we use the perturbation method
to get an approximate  bounce solution when the initial charge
density is very small.

Once we find the $O(3)$ bounce solution $f_b, \eta_b$, we can
calculate its action $S_E + \sigma$ from Eqs. (3.2) and (3.5). By
using Eqs.(3.7) and (3.8), we can see the combined  action becomes
$$ S_E  + \sigma = \int d^4x \left\{ {1\over 2} (\partial_l  f_b)^2 +
{1\over 2} f^2_b (\partial_l  \eta_b)^2 + U(f_b)   \right\}
\eqno\eq  $$
The bubble nucleation per unit  volume is then given by
$Ke^{-B}$ where
$$ B = (S_E + \sigma)({\rm bounce}) - (S_E + \sigma)({\rm background})
\eqno\eq $$
While the action for the bounce $S_E +\sigma$ could be infinite, the
difference $B$ between that of the bounce and that of the initial
configuration should be finite when one expects a finite tunneling
rate. While we will not attempt to calculate the effect of the
fluctuations around the bounce solution, we note that the field
fluctuations $\delta f, \delta \theta$ should be kept  real in the
functional integral. This is exactly what happens in a gaussian
integral
$$ \int dx e^{-x^2 + ipx} $$

To find out the escape point or the bubble configuration at the
nucleation moment, we use the
time translation and reflection symmetries of the action under $\tau
\rightarrow -\tau$ and $\eta \rightarrow -\eta$ of $S_E + \sigma$ to
choose the origin to be the center of the bounce so that
$$  \eqalign{ &\  {\partial f_b \over \partial \tau} (\vec{r},\tau=0 ) = 0 \cr
&\ \eta_b(\vec{r},\tau=0) = 0 \cr}
\eqno\eq $$
As in our problem charge density remains real in Minkowski and
Euclidean times, it is natural to identify the initial charge density
of the bubble to be given by that of the bounce. The initial bubble
configuration is then
$$ \eqalign{ &\   f( \vec{r},t=0 ) = f_b( \vec{r},\tau =0 ) \cr
 &\  \theta(\vec{r},t  = 0) = 0  \cr
&\   {\partial f \over \partial t} ( \vec{r}, t =0 ) = 0 \cr
&\    f^2 {\partial \theta \over \partial t} ( \vec{r}, t=0) =
     f^2_b { \partial \eta_b \over \partial \tau}(\vec{r},\tau=0)  \cr}
\eqno\eq $$
In usual quantum tunneling, momenta are imaginary and coordinates are
real under the potential barrier, and we find the escape point in the
coordinate space. In our case, the charge or momentum density and the
angle variable have changed their role.  Under the barrier the angle
variable is imaginary and the charge density is real, and we find the
escape point in the charge density.

Once we know the initial bubble configuration (3.13), we can solve the
field equation in the Minkowski time to find out how a given bubble
evolve. The bounce solution $(f_b, \eta_b)$ can also be analytically
continued to a solution in Minkowski time,
$$ \eqalign{ &\ f( \vec{r},t) =   f_b( \vec{r},it)  \cr
	&\ \theta(\vec{r},t) =- i\eta_b(\vec{r},it) 	\cr}
\eqno\eq $$

A further insight about bubble nucleation with global charge can be
gained by using the dual formulation.\refmark\rKlee It is well known
that Goldstone bosons can be described by an  antisymmetric tensor
field, $B_{\mu\nu}$.  In the Minkowski time, the dual lagrangian is
given by

$$ {\cal L}_{DM} = {1\over 2}(\partial_\mu f)^2 + {1\over 12 f^2
}H_{\mu\nu\rho}^2 - U(f)
\eqno\eq
$$
where $H_{\mu\nu\rho} = \partial_\mu B_{\nu\rho} + \partial_\nu
B_{\rho\mu} + \partial_\rho B_{\mu\nu} $. The field strength of the
antisymmetric tensor is related to the original current by
$$ f^2 \partial^\mu \theta = {1\over 6}  \epsilon^{\mu\nu\rho\sigma}
H_{\mu\nu\rho}
\eqno\eq $$
The uniform initial charge density becomes the condition of the
uniform `magnetic' field $H_{123}$.  In Euclidean time, there will be
no boundary term arising from the wave function $\Psi(f, B_{ij})$. The
Euclidean lagrangian  becomes
$$ {\cal L}_{DE} = {1\over 2} (\partial_\mu f)^2 + {1\over 12 f^2}
H_{\mu\nu\rho}^2 + U(f)
\eqno\eq $$
The bounce equation becomes
$$ \eqalign{
&\ \partial_\mu^2 f + {1\over  6f^3} H_{\mu\nu\rho}^2 - U'(f) = 0 \cr
&\ \partial^\mu ( {1\over f^2} H_{\mu\nu\rho}) = 0 \cr}
\eqno\eq $$
We can relate the Euclidean fields either through the Euclidean time
dual transformation or by comparing the current. The relation between
the antisymmetric tensor field and the angle variable in the Euclidean
time is given by
$$ f^2 \partial^\mu \eta =  {1\over 6}  \epsilon^{\mu\nu\rho\sigma}
H_{\nu\rho\sigma}
\eqno\eq $$
Thus, the bounce solution in terms of the $B_{\mu\nu}$ field would be
real and there would be no contribution to the bounce action from the
boundary.

\chapter{Examples}

We are now in position to examine in more detail three cases of the
phase transition which we have discussed in Sec.2. To be more specific
we choose the potential to be given by Eq.(2.12).  {\it Case A} has
the initial configuration in the symmetric phase. The potential energy
has the absolute minimum at the asymmetric phase.  For {\it Cases B,
C} the potential energy $U(f)$ has the absolute minimum at the
symmetric phase and the local minimum at the asymmetric phase. The
initial configuration of {\it Case B} is at the asymmetric phase and
the initial configuration of {\it Case C} is at the symmetric phase.
In {\it Cases A, B} the tunneling would proceed even when there is no
initial charge density because the initial configurations are at the
metastable points of $U(f)$. Introducing a small amount of
charge density would not change much of the original bubble nucleation.
Thus we would expand perturbatively the bounce solution by the initial
charge density and see how the zeroth order $O(4)$ symmetric solution
deforms.  These are the cases we will examine closely in this
section. For {\it Case C}, we do not have the zeroth order bounce
solution because the quantum tunneling occurs solely due to the charge
density. However, we can still get some insight for this case as we
will see  later.

Let us consider first the cases where the initial configuration is
unstable even without any charge density. The symmetric phase of the
potential $U_1$ and the asymmetric phase of the potential $U_2$ in Fig.2
are such initial configurations.  When there is no initial charge
density, the bounce solution can be obtained by the O(4) symmetric
ansatz, $\tilde{f}(s \equiv \sqrt{\vec{r}\,^2 + \tau^2})$.\refmark\rColeman
Let us assume that the thin wall approximation works.  We call that
inside the wall, $f=f_i$ and outside the wall $f=f_e$. The wall radius
$a$ can be determined as follows.  Suppose that the potential energy
difference $\Delta U = U(f_e) - U(f_i)>0 $ between $f_e$ and $f_i$
phase is small.  Then the bubble radius will be large and we can
approximate the bubble wall as a domain wall separating two phases.
This wall satisfies the equation, $ \partial_x^2 f + U'(f) = 0 $,
neglecting the potential energy difference. Define the tension of the
wall to be the action density per unit three volume, $T = \int dx [
(df/dx)^2 /2 + U(f)]$.  The gain of the action due to this true vacuum
bubble of the radius $a$ is then
$$ S(a) = 2\pi^2 T a^3 - {\pi^2 \over 2} \Delta U a^4  \eqno\eq $$
At $a= 3T/\Delta U$,  $S(a)$ takes the maximum value $  27\pi^2
T^4/2(\Delta U)^3$, which is the $B$ factor in the bubble nucleation rate.

We ask what happens to this $O(4)$ symmetric thin-wall bounce solution
if we introduce a small initial charge density. From Eq.(3.7), we see
that the equation of the angle variable is the first order in the charge
density and the $f$ field equation has a second order correction to
the bounce equation of the zero charge density.  Thus, we can solve
Eq.(3.7) by a perturbative expansion around this $O(4)$ symmetric
background. The phase variable will be first order in charge density
and satisfies  the current conservation
$$ \partial_l (\tilde{f}^2(s)\partial_l\eta) = 0    \eqno\eq $$
with the boundary condition  $f_e^2 \eta(\vec{r}, \tau = \pm \infty) =
\rho_0$ with the initial charge density $\rho_0$.
The above equation can be interpreted as a boundary problem of a
dielectric media in four space dimensions.  The electric field is $
\partial_\mu \eta$ with the potential $\eta$ and the electric
displacement is $J_\mu = f^2 \partial_\mu \eta$ with the dielectric
constant $f^2$. Eq.(4.2) implies the boundary condition at the wall
that the normal component of $J_\mu$ and the tangential component of
$\partial_\mu \eta $ should be continuous.  The boundary condition at
infinity is that there is a constant external electric displacement
field $J_\tau = \rho_0$. For a given $O(4)$ symmetric configuration
described before, it is trivial to find the potential $\eta_{e,i}$
outside and inside the thin wall,
$$ \eqalign{ &\ \eta_e =
 {\rho_0 \over f_e^2}\left( 1  +{f_e^2 - f_i^2 \over 3f_e^2 + f_i^2}
 {a^4 \over s^4 }\right) \tau  \cr
&\ \eta_i = {\rho_0 \over f_e^2} \left( {4f_e^2 \over 3f_e^2 + f_i^2}
\right) \tau   \cr}
\eqno\eq  $$
The charge density $J_\tau$ at the $\tau = 0$ would be the charge
profile of the bubble at the moment of nucleation as shown in
Eq.(3.13).  Let us now examine the implications of this solution (4.3)
in various cases.  The correction to the $f$ field would be second
order and will be considered in each case.

\noindent{ \it  Case A: from the symmetric phase to the asymmetric phase}

Let us first consider the case when the asymmetric phase is the ground
state and the initial configuration is near the symmetric phase. Thus,
$f_e \sim 0 << f_i \sim v $. Since there is no initial charge density,
the previous argument would imply that $f_e =0$.  When we introduce
uniform charge density in the initial configuration, the initial value
of $f$ would be given by Eq.(2.13) with $\theta = mt$, invalidating
our assumption $\tilde{f} \sim {\cal O}(1)$.  Here let us assume simply that
 $f_e$ is nonzero  even when there is no charge, say
due to a small bump in the potential at $f=0$. This would not change
the physics of tunneling under consideration much and allows us to use
Eq.(4.3).

\FIG\ffour{The  bounce solution for {\it Case A}. The dashed circle
is the wall of the $O(4)$ symmetric bounce solution.  The dotted lines
indicate the charge flow. The solid ellipse is the wall of the
deformed bounce.  }

The global current $J_\mu$ around the $O(4)$ symmetric bounce solution
can be obtained from Eq.(4.3). Outside the thin wall $(\zeta>a)$,
$J_\mu = f_e^2 \partial_\mu \eta_e$ and inside the thin wall
$(\zeta<a)$, $J_\mu = f^i_2 \partial_\mu \eta_i$. The energy per
charge inside the wall is small and so the charge is attracted to the
interior region, making the charge density inside the bubble be higher
than that outside.  The charge density profile of the bubble at the
moment of nucleation would be given by $J_\tau $ at $\tau = 0 $,
$$ \eqalign{ &\ J^0_{exterior} =  (1 - {a^4 \over r^4}) \rho_0 \cr
&\ J^0_{interior} = 4\rho_0 \cr}
\eqno\eq $$
when $ f_i >> f_e $. The charge density inside the bubble is {\it
four} times larger than the initial charge density.  From Eq.(4.4), we
can see $\int d^3x ( J^0 - \rho_0) = 0 $, implying that the charge
inside the bubble came from the region near the bubble wall.  Fig.4
shows this global current on the $O(4)$ symmetric solution background.

Let us now consider the effect of the charge flow on the $f$ field.
The $f$ field equation (3.7) can be expanded around the $O(4)$
symmetric solution  as
$$ \partial_\mu^2 \delta f  -
U''(f)\delta f =  - { J_\mu^2 \over \tilde{f}^3}
\eqno\eq $$
where $J_\mu^2$ can be obtained from Eq.(4.3) and $\tilde{f} $ would
be given by the $O(4)$ symmetric solution.  Rather than try to solve
this partial differential equation, let us approach the problem more
qualitatively. When $f_e<< f_i$, one can show that at the north and
south poles $J_\mu^2 \sim 16
\rho_0^2$ both inside and outside and that at the equator $J_\mu^2 \sim
0 $ outside and $J_\mu^2 \sim 16 \rho_0^2$ inside. The centrifugal
term $J_\mu^2/2f^2$ would be then important at the poles but not at
the equator.  Directly from the $f$ equation (3.7) and the previous
argument about the thin-wall approximation, we see the centrifugal
term reduces the tension on the domain wall by reducing the effective
potential energy barrier.  Since the tensions at the two poles are
lower while the tensor at the equator remain constant, the curvature
at poles would be larger than $1/a$ and that at the equator will
remain $1/a$ where $a= 3T /\Delta U$ is the radius of the $O(4)$
symmetric shell.  ( In addition $\Delta U$ in increased at the pole
and remain unchanged at the equator, amplifying the curvature change
between the poles and the equator.)  Consequently, the O(4) symmetric
wall would be shrunk at the equator.  We take a liberty to choose this
wall configuration to be an ellipsoid
$$ {r^2 \over b^2 } + {\tau^2 \over ab  }   = 1   \eqno\eq $$
where  $b<a$ takes a complicated function of $\rho_0$.
Fig.4 shows the deformed bounce wall.

Let us now think about the bubble nucleation and evolution. The bubble
of true vacuum will nucleate with a radius $b$ and a higher charge
density $4\rho_0$  and then expand. As argued in Eq.(3.14), we
can analytically continue the bounce solution to find out how the
bubble will evolve. From Eq.(4.6), we see that the bubble wall
trajectory will be given by $ R(t) = \sqrt{{b\over a}} \sqrt{ t^2 + ab
}$.  Since $b<a$, the terminal velocity $v_t = \sqrt{b/a}$ is less
than 1.

How do we understand the finite terminal velocity? Energy conservation
implies that the change of the wall energy comes from  the
potential energy difference,
$$ d\left({ 4\pi  T(R) R^2 \over \sqrt{1-\dot{R}^2} } \right)
= d\left( {4\pi \over 3} \Delta U R^3 \right)   \eqno\eq $$
The tension of the bubble wall surface could depend on $R$.  Integrating
Eq.(4.7), we get
$$ \dot{R} = \sqrt{ 1 -  {3T(R) \over \Delta UR } }
\eqno\eq $$
When there is no charge density, $T, \Delta U$ are fixed and we see
that the terminal velocity is the light speed. If the tension grows
linearly with the radius for large $R$, $T(R) \sim \alpha R$, the
terminal velocity $v_t =\sqrt{b/a}$ will be
$$ v_t = \sqrt{ 1- {3\alpha \over \Delta U}} < 1 \eqno\eq $$
The growth of tension, or the energy density of wall per unit area,
can be understood by considering the phase variable $\theta$. While
the charge density $f_i^2 \dot{\theta}$ inside the bubble is larger
than the charge density $f_e^2\dot{\theta}$ outside, it is but not
large enough to keep the phase variable space independent since $f_i>>
f_e$. The phase increases by $mt$ outside the bubble and more slowly
inside the bubble, leading to the increasing its space gradient at the
bubble wall. This is what we think is the source of the increasing
tension or energy density at the bubble wall.

Let us now remind ourselves that the sound speed (2.15) in the
symmetric phase is much smaller than the terminal speed (4.9) when the
initial charge density is small. Thus, the bubble wall forms a sort of
supersonic front in the symmetric phase. However, our analysis is not
accurate enough to compare the terminal speed and the sound speed
(2.17) at the inside asymmetric phase.  ( When the initial charge
density is large enough, there is a possibility that the terminal
speed is less than the sound speed in the symmetric phase. )  The thin
wall approximation would fail eventually, because there is not enough
charge lying outside the expanding bubble to keep the charge density
inside the bubble to be four time larger than the initial value. The
charge density profile around the expanding bubble should become more
smoothly changing.

\noindent{ \it Case B: from the asymmetric phase to the symmetric phase }

Let us now consider the case where the initial state is the broken
phase which has the higher potential energy than that of the symmetric
phase.  When there is no initial charge density, there will be an O(4)
symmetric bounce solution, inside which the scalar field takes value
near the symmetric phase, $f_i<< f_e$.  Again we ask what is the
consequence of the small initial charge density. Eq.(4.3) implies  how
the current flows around this bounce solution.  The charge density at
the moment of bubble nucleation would be given by
$$ \eqalign{ &\  J^0_{exterior} = (1 + {a^4 \over 3r^4}) \rho_0 \cr
&\ J^0_{interior} \approx  0 \cr}
\eqno\eq$$
when $f_i<< f_e$.  The charge is excluded from the symmetric vacuum
region. Fig.5 shows the charge flow around the $O(4)$ symmetric bounce
solution.

\FIG\ffive{The  bounce solution for {\it Case B}. The dashed circle
is the wall of the $O(4)$ symmetric bounce solution.  The dotted lines
indicate the charge flow. The solid ellipse is the wall of the
deformed bounce.  }

The effect of the charge flow on the $f$ field is given by the
centrifugal term $J_\mu^2 /2f^2$ as in Eq.(4.5).  We can calculate
$J_\mu^2$ for our bounce solution. Since $f_e>> f_i$, Eq.(4.3) implies
that at north and south poles $J_\mu^2 \sim 0$ both inside and outside
and that at the equator $J_\mu^2 \approx 0 $ inside and $J_\mu^2
\approx {16 \rho_0^2 \over 9}$ outside.  This raises the energy
density of the false vacuum at the equator, lowering the barrier
energy and increasing $\Delta U$. This in turn lowers the tension of
wall at the equator. The tension at the poles would remain unchanged
and so the bounce solution would be elongated at the equator. Now for
the sake of the argument, we again approximate the wall as an
ellipsoid,
$$   { r^2 \over ab} + {\tau^2 \over b^2}  = 1   \eqno\eq  $$
where $b<a$ is a complicated function of $\rho_e$.
The deformed bounce solution is shown in Fig.5.

In the Minkowski time, the bubble wall trajectory is given by $R(t)=
\sqrt{a/b}  \sqrt{t^2 + b^2}$, with the terminal velocity
$v_t = \sqrt{a/b} > 1$, which clearly violates the causality.
Something should happens before the wall speed becomes the light
speed.  When there is a tachyonic mode, we say there is an unstable or
exponentially growing mode.  There are many possibilities.  Since
charge is pushed out from the bubble and is accumulated at the wall,
the wall could stop expanding. Or the thin wall approximation could
break down before the wall reachs the light speed.

In {\it Case B}, the charge is pushed way from the bubbles and
accumulated at the initial asymmetric phase. At the end of phase
transition, we would be left with the islands of the original phase
with high charge density, which are exactly Q-balls floating in the
symmetric phase.

\noindent{\it Case C: forming Q-balls by quantum tunneling}

If the minimum of $\sqrt{2U/f^2}$ is lower than the mass of the
charged particles in the symmetric phase, charge likes to clump to
Q-balls. Suppose the initial configuration lies at the symmetric phase
with very small charge density and is stable under local fluctuation.
Since the minimum of the potential is chosen to be the symmetric
phase, the initial configuration would be stable if there is no charge
density. After small uniform charge density is introduced, the initial
configuration however becomes unstable under the quantum mechanical
tunneling transition to form Q-balls. Since we start from the minimum
of the potential, we do not have the bounce solution at zero charge
density.  However, we can still gain some understanding of the general
features of the bounce solution from what we learned in {\it Case A}.

First, the bubble at the moment of nucleation would be a Q-Ball of the
minimum size, where the surface energy is as important as the volume
energy.  The charge density $\rho_*$ inside the Q-ball would be much
larger than the initial charge density, $\rho_0$.  The minimum size of
a Q-ball could be rather large if $w_*$ is very close to $m$ so that
the energy gain by the charge is small and so a lot of charge is
needed to compensate the surface energy. We know that there will be a
large current will flow into inside the bounce wall from outside in
this case because the interior charge density $\rho_i = f_i^2 w_*^2$
is much larger than $\rho_0$.  With a similar argment given to {\it
Case A}, the current would be large inside and outside wall at the
poles, and would be zero outside and large inside the equator. The
energy density outside the bounce wall at the pole is larger than that
outside the bounce wall at the equator. The barrier energy at the
poles would in turn be lower than that at the equator, and so the wall
tension at pole will be lower than that at the equator.  Thus, the
bounce solution in {\it Case C} would also be elongated along the
$\tau$ direction in {\it Case A}. Fig.6 shows such a bounce solution
for  Q-ball nucleation.

Once a Q-ball is nucleated, it will grow but very slowly. The reason
is that a Q-ball can grow only when it swallows the charge from
outside and that there is not much charge around it because the
formation itself have already diluted the initial charge density
around its neighborhood. This is consistent with the picture that the
bounce solution is elongated along the $\tau$ direction, which also
implies a slow terminal velocity as argued after Eq.(4.9). The
explicit nature of the Q-ball nucleation and expansion would however
require the better analysis and would not be attempted here.

\FIG\fsix{The  bounce solution for {\it Case C}.   The dotted lines
indicate the charge flow. The solid ellipse is the wall of the
bounce for Q-ball nucleation.  }

\chapter{ Discussion}

We have studied the phase transitions in the theory of a complex
scalar field with a global U(1) symmetry when there is non zero
initial charge density. We have discussed the metastability condition
on the possible initial configurations and the possible inhomogeneous
final configurations. We argued that there are many cases  of the phase
transitions to be studied in the theory.  We have set the Euclidean
formalism of the bubble nucleation when there is nonzero charge
density. We applied our formalism to the case when the initial charge
density is small and when the phase transitions involve the symmetric
phase as the initial configuration or a part of the finial
configuration. Here we studied the characteristics of the bounce
solutions and the bubble evolution. Our system is shown to have a rich
variety of the possible phase transitions, and could be a good simple
toy model of the phase transition involving charges, the supersonic
bubble wall,  and the sound wave.

However, there are still many loose ends and questions we have not
attempt here.  One of the interesting questions seems what is the
later development of the bubbles. Depending on the cases of the phase
transitions, there is a possibility of  rich dynamics.  Additional
interesting questions to be explored are about the phase transition
between the asymmetric phases and about the phase transition when the
initial charge density is not small.  Finally, we note that it is
rather straightforward to extend our formalism in Sec.3 to the case
involving the nonzero local gauge charges.

\vskip 1in

\centerline{\bf Acknowledgement}

This work is in part supported by the NSF under Grant No.
PHY-92-45317, Department of Energy, the NSF Presidential Young
Investigator program and the Alfred P. Sloan Foundation. The author
likes to thank colleagues at Institute for Advanced Studies, Institute
for Theoretical Physics and Aspen Physics Center, where a part of this
work was  done.  The author is also grateful to S. Coleman, K. Freese
and L. Widrow for insightful discussions.

\vfill \endpage
\refout

\vfill \endpage
\figout
\end